\documentclass{raa}            

\usepackage{graphicx,times}             
\usepackage{natbib}
\usepackage{amssymb,amsmath}
\bibpunct{(}{)}{;}{a}{}{,}

\usepackage[pagebackref=true]{hyperref}

\begin{document}

  \title{Environmental Dependence of Star Formation and Galaxy Colors around Abell 2029}

   \volnopage{Vol.0 (20xx) No.0, 000--000}      
   \setcounter{page}{1}          

   \author{Xiaolan Hou 
      \inst{1}
   \and Heng Yu
      \inst{1}
   }

   \institute{
             School of Physics and Astronomy,Beijing Normal University, Beijing, 100875, China \\{\it email: yuheng@bnu.edu.cn}\\
\vs\no
   {\small Received 20xx month day; accepted 20xx month day}}

\abstract{Environmental processes drive galaxy evolution, with the impact varying significantly across different stellar masses. We present a comprehensive environmental analysis of the galaxies within a $10^\circ \times 10^\circ$ field around A2029, utilizing high-density spectroscopic data from the DESI and SDSS surveys. We investigate the quenched fraction ($f_Q$) and red fraction ($f_{red}$) as functions of local surface density ($\log_{10} \Sigma_5$) across three stellar mass intervals (low-mass: $9.5 \le \log M_\star/M_\odot < 10.0$; medium-mass: $10.0 \le \log M_\star/M_\odot < 10.5$; high-mass: ($\log M_\star/M_\odot \geq 10.5$ ). Our results show that, for galaxies of all masses, both star formation activity and galaxy color are strongly correlated with the local density.
Although the environmental dependence of both the quenched and red fractions is somewhat weaker in low-mass galaxies than in their high-mass counterparts, the variations remain significant. This suggests that galaxy colors, even for low-mass systems, can serve as effective tracers of large-scale structure.
\keywords{galaxies: clusters: individual: Abell 2029 --- galaxies: evolution --- galaxies: star formation}
}

   \authorrunning{X.-L. Hou, \& H. Yu}            
   \titlerunning{Environmental Dependence of Star Formation and Galaxy Colors around Abell 2029}  

   \maketitle

%
%
\section{Introduction}           
\label{sect:intro}

Galaxy clusters are the most massive gravitationally bound structures in the universe, serving as the primary nodes within the cosmic web \citep{Springel2005, Kravtsov2012}. These systems consist of hundreds to thousands of galaxies embedded within expansive dark matter halos, and are characterized by a hot, ionized intracluster medium (ICM) that accounts for a significant fraction of the total cluster mass \citep{Allen2011}. Within the framework of hierarchical structure formation, clusters grow through the continuous accretion of small galaxy groups and the flow of matter along large-scale filaments \citep{Bond1996, AragonCalvo2010}.

As galaxy clusters evolve, the properties of the galaxies within them also undergo corresponding changes. Under the combined influence of mass and environment, galaxies transition from blue star-forming galaxies to red, quenched galaxies\citep{Strateva2001, Baldry2004, Peng2010, Peng2015, Bianconi2021}. In sufficiently massive galaxies, feedback from its own active galactic nucleus (AGN) rapidly depletes the internal cold gas and prevents further inflow of external gas.
Upon entering the dense region of cluster core, galaxies experience rapid quenching triggered by mechanisms such as ram-pressure stripping\citep{Gunn1972, Steinhause2016, Boselli2022}, frequent galactic harassment\citep{Boselli2014, Bahe2015}, or starvation\citep{Zinger2018, Maier2019}.

Even in lower-density regions, such as the outskirts of galaxy clusters and filaments, star formation within galaxies is subject to a slight but persistent environmental influence, often referred to as 'pre-processing' \citep{Fujita2004, McGee2009, Sarron2019}. 
Observational evidence consistently shows that galaxies associated with these infalling groups and filaments exhibit higher quenched fractions compared to their counterparts in truly isolated field environments \citep{Porter2008, Kraljic2018}. 
By comparing galaxy properties across different environments, we can quantify the impact of the environment on galactic evolution. However, due to observational limitations, it is challenging to reliably identify low-density regions, such as filaments\citep{Libeskind2018, Rost2020}. Nevertheless, recent galaxy redshift surveys, such as the Dark Energy Spectroscopic Instrument \citep[DESI]{DESI2025}, enable preliminary explorations at low redshifts.

Abell 2029 (A2029; z=0.0787) is one of the most massive cool-core clusters in the local universe. While traditionally viewed as a dynamically relaxed system, deep X-ray observations have revealed complex internal features, such as sloshing spirals and merger shocks, indicating that the cluster is currently undergoing active growth and assembly \citep{Dressler1978, Clarke2004, Paterno-Mahler2013, Sohn2019a}.

On a larger scale , A2029 serves as the gravitational hub of an expansive supercluster complex. Comprehensive spectroscopic surveys have identified over 1,200 member galaxies within this system, including significant infalling subsystems such as Abell 2033 (A2033) to the north and the Southern Infalling Group (SIG) \citep{Sohn2019a}. These sub-clusters are linked to A2029 by a large-scale gaseous bridge, which has been identified as a cosmic filament ($\sim$ 1 Mpc wide and 20 Mpc long) tracing the warm-hot intergalactic medium \citep[WHIM;][]{Walker2012, Nicastro2018}. The hierarchical arrangement of these structures, mapped using advanced algorithms like the Blooming Tree (BT) method, illustrates how the A2029 system continues to assemble from the surrounding cosmic web \citep{Yu2025}.

The complex environment surrounding A2029 offers an ideal laboratory for investigating the impact of external physical processes on galaxy evolution.
In this study, we utilize high-density spectroscopic data from the DESI survey and the Sloan Digital Sky Survey (SDSS) to perform a detailed analysis of galaxies across varying stellar mass regimes aroud A2029. In Section \ref{sec:data}, we describe the datasets and the sample selection criteria. Section \ref{sec:res} presents our analysis and results. Finally, we summarize in Section \ref{sec:con}.

\section{Data and Sample} \label{sec:data}

We define the primary survey area as a $10^\circ \times 10^\circ$ field centered on A2029 ($\mathrm{RA} = 228.0^\circ, \mathrm{Dec} = 5.5^\circ$). At the cluster's redshift, this field corresponds to a physical scale of approximately $53 \times 53 \text{ Mpc}^2$. This wide-field coverage extends to a cluster-centric radius of $\sim 14R_{200}$ (where $R_{200} = 1.91 \text{ Mpc}$), enabling a comprehensive investigation of the cluster, its surrounding groups, filaments, and infalling field galaxies \citep{Sohn2019a, Sohn2019b, Yu2025}. 

\subsection{Data Completeness}

The spectroscopic redshifts used in this study are primarily from the DESI Data Release 1 \citep[DR1;][]{DESI2025}. To maximize sampling density and mitigate the impact of fiber collisions in high-density regions, we supplement the DESI data with spectroscopic redshifts from SDSS DR18 \citep{Almeida2023}.
For galaxies observed by both DESI and SDSS, we prioritize the DESI redshift due to their generally higher signal-to-noise ratios. This procedure yields a refined spectroscopic catalog containing 120,313 unique galaxies with high-quality redshifts within the $10^\circ \times 10^\circ$ field. Among these, 109,547 galaxies are from DESI, while 10,766 are contributed by SDSS.
The star formation rate ($\mathrm{SFR}$) and stellar mass ($M_\star$) measurements are sourced from \cite{Zou2024} for DESI galaxies, where both quantities were derived via spectral energy distribution (SED) fitting using the CIGALE code \citep{Boquien2019}.
And from the SDSS galSpecExtra tables \citep{Brinchmann2004} for SDSS objects.

To estimate the completeness of the spectroscopic data, we selected a photometric sample from the DESI Legacy Surveys Data Release 9 (LS DR9) sweep catalogs \citep{Dey2019}. The Legacy Surveys provide deep, three-band ($g, r, z$) imaging that serves as the basis for target selection. 
For each galaxy, we calculate the $r$-band magnitude as: $r = -2.5 \log_{10}(F/ MW_{TRANS}) + 22.5$ where $F$ represents the $r$-band flux and $MW_{TRANS}$ accounts for the Galactic extinction correction. 
This photometric catalog serves as the parent sample from which our spectroscopic targets are drawn. 
The sample includes 1,025,686 extended objects within the magnitude range $13.5 \leq r \leq 22$ mag.

\begin{figure}[h]
    \centering
    \begin{subfigure}[b]{0.7\textwidth} 
        \centering
        \includegraphics[width=\linewidth]{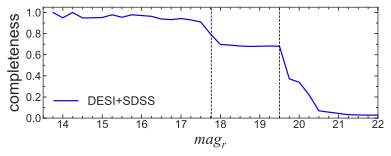}
    \end{subfigure}   
    \vspace{0.5cm} 
    \begin{subfigure}[b]{0.7\textwidth} 
            \centering
        \includegraphics[width=\linewidth]{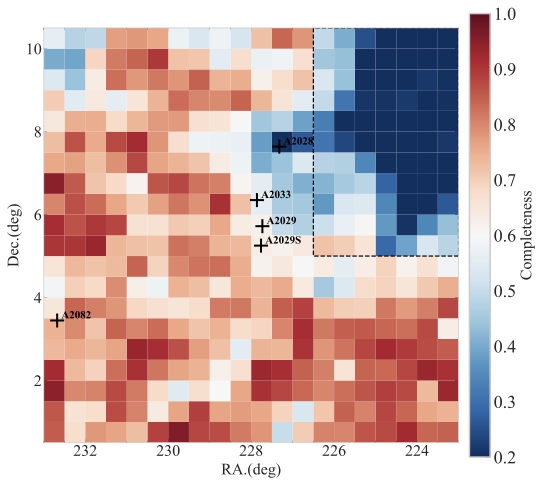}
    \end{subfigure}
    \caption{Spectroscopic data completeness. \textit{Upper panel}: Spectroscopic completeness as a function of \textit{r}-band magnitude. \textit{Bottom panel}: Spatial distribution of the spectroscopic completeness. The colorbar represents the completeness ratio.}
    \label{fig:comp}
\end{figure}

The spectroscopic completeness, defined as the ratio of galaxies with confirmed redshifts to the total photometric parent sample, is illustrated in Fig. \ref{fig:comp}. 
Based on the spatial completeness distribution over a 20 $\times$ 20-pixel grid,
We found that the data coverage in the northwest portion of the field of view is significantly incomplete.
This is due to the non-uniform sky coverage of the DESI DR1.
To prevent this localized deficit from biasing our global completeness statistics and density measurements, we apply a spatial mask ($\mathrm{RA} < 226.5^\circ$ and $\mathrm{Dec} > 5.0^\circ$) to exclude this area from the magnitude completeness profile shown in Fig. \ref{fig:comp} and all subsequent analysis.

From the magnitude completeness profile, we identified a clear decline in completeness beyond $r = 19.5$. This corresponds to the flux limit of the DESI Bright Galaxy Survey (BGS) bright sample ($r < 19.5$) \citep{Hahn2023}.
Consequently, we impose a magnitude cut of $r \leq 19.5$ for all subsequent analysis. This threshold ensures a spectroscopic completeness of approximately 79\% within our study area (64,385 galaxies with spectroscopic data and 81,544 galaxies with photometric data). 

We examined the impact of this fiber-collision effect on spectroscopic completeness.
The DESI fiber housing has a physical positioner boundary of approximately 6 mm, corresponding to an angular scale of about $85''$ (\citealt{Silber2023}). Among the 17,159 galaxies lacking spectroscopic observations, only 1,566 have no neighboring galaxies within a radius of $85''$. In other words, approximately 91\% of galaxies without redshift measurements can be attributed to the fiber-collision effect.
Given that spectroscopic incompleteness primarily affects high-density environments \citep{Ellison2008}, where fiber collisions are more severe, the extremely small number of unobserved isolated galaxies indicates that this hardware-induced suppression is strongly confined to the collision scale. Therefore, although local densities may be systematically underestimated on small physical scales ($85''$ corresponds to $\sim125$ kpc at $z=0.078$), this localized missing-data effect is unlikely to introduce significant bias in environmental density estimates on larger scales.

\subsection{Sample Properties}
\begin{figure}
\centering
\includegraphics[width=0.7\textwidth]{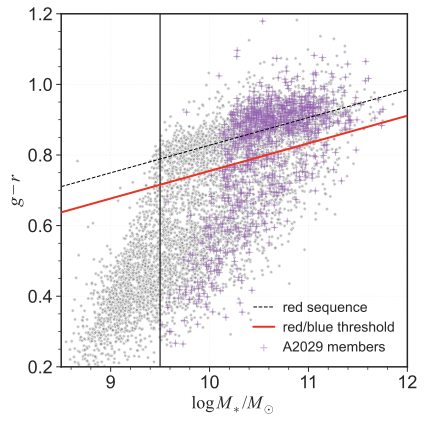}
\caption{Distribution of galaxies in the $(g-r)$ color versus stellar mass ($\log M_\star/M_\odot $) plane. Purple crosses represent the member galaxies of the A2029 cluster as identified by the BT algorithm \citep{Yu2025}. The diagonal solid black line denotes the best-fit red sequence, defined by the relation $g-r = 0.074 \ \log M_\star/M_\odot + 0.095$. The parallel red line indicates the $1\sigma$ deviation from this fit. Galaxies located above this $1\sigma$ threshold are classified as red, while those below are categorized as blue. The vertical solid black line marks the 90\% stellar mass completeness limit at $\log M_\star/M_\odot = 9.5$.}
\label{fig:M_color}
\end{figure}

To extract a representative sample of galaxies associated with the A2029, we restrict our analysis to the redshift range $0.06 \le z \le 0.09$. 
The color-magnitude relation of these galaxies is shown in Figure \ref{fig:M_color}, revealing a clear bimodal distribution is visible. This distribution further validates the adopted mass completeness limit. We fitted the member galaxies (red crosses) of the A2029 cluster as identified by the BT algorithm \citep{Yu2025} which is based on SDSS spectroscopic data. The derived red sequence is $g-r = 0.074 \log(M_\star/M_\odot) + 0.095$, represented by the solid black line in Figure \ref{fig:M_color}.
Galaxies in our sample are classified as red or blue based on the $1\sigma$ dispersion of the red sequence.
Specifically, galaxies located above this boundary are classified as red, while those below are categorized as blue.

To construct a mass complete sample, we followed a similar methodology of \cite{Pozzetti2010}, defining the mass completeness limit as the 90th percentile of the limiting stellar masses (the mass each galaxy would have at the survey flux limit) of the 20\% faintest galaxies.
Given our magnitude limit of $r = 19.5$, the stellar mass completeness limit is estimated to be $\log M_\star/M_\odot = 9.5$. Adopting this threshold mitigates, we mitigate potential selection biases arising from the exclusion of low-luminosity galaxies. 
The final spectroscopic sample comprises 4,550 galaxies within the defined $10^\circ \times 10^\circ$ field, of which 3,749 galaxies are from DESI and 801 from SDSS.

\begin{figure}
\centering
\includegraphics[width=0.7\textwidth]{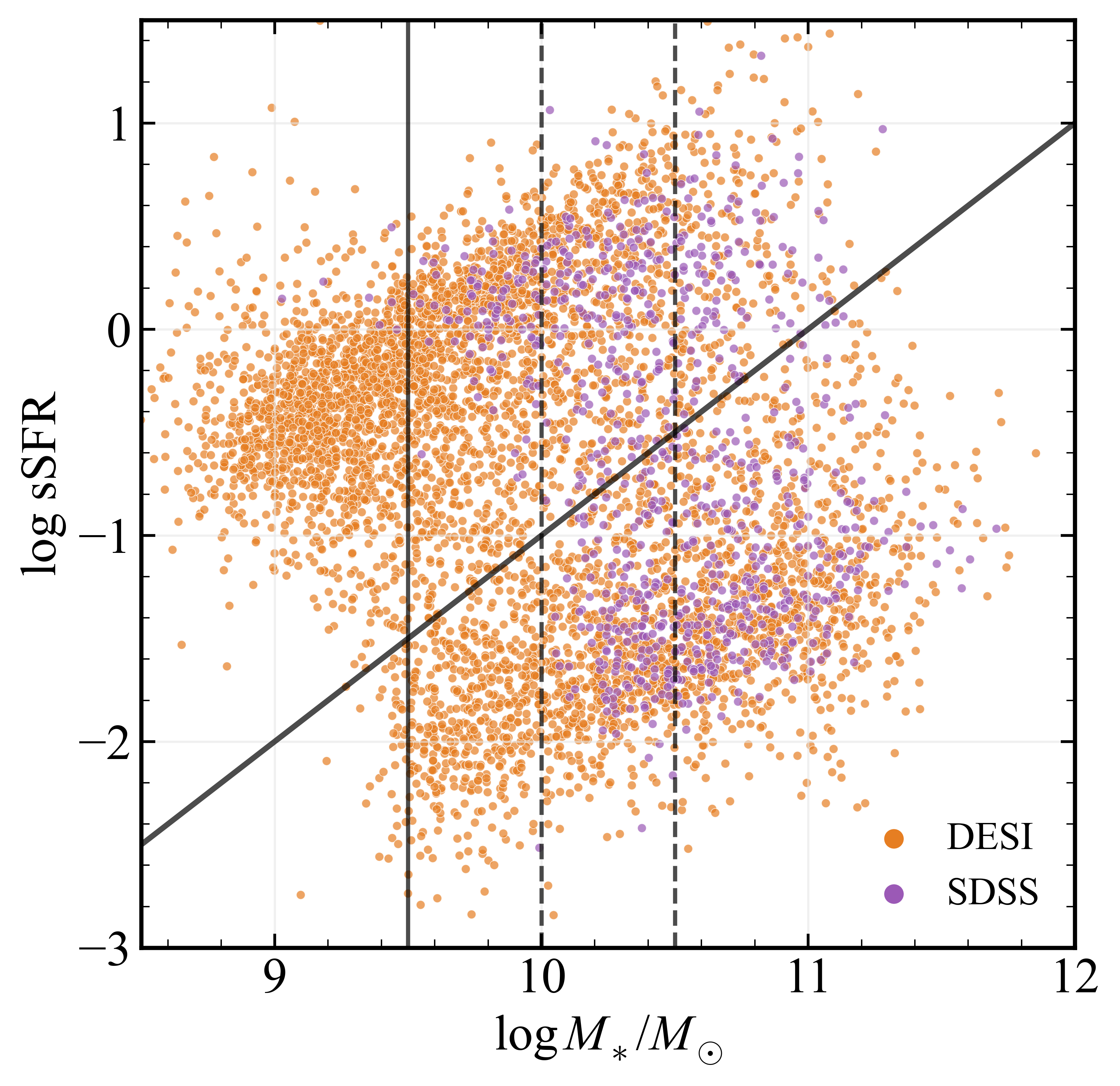}
\caption{Distribution of galaxies in the $\log \mathrm{SFR}$ versus $\log M_\star/M_\odot $ plane. Purple and orange points represent galaxies with spectroscopic redshifts from the SDSS and DESI surveys, respectively. The diagonal black line corresponds to a constant specific star formation rate of $\log \mathrm{sSFR} = -11 \text{ yr}^{-1}$, which serves as the empirical threshold for separating the star-forming main sequence from the quenched population. The vertical solid line is same as Fig.\ref{fig:M_color}, the vertial dashed lines are $\log M_\star/M_\odot =10$ and $\log M_\star/M_\odot =10.5$ .}
\label{fig:M_SFR}
\end{figure}

In addition to color, we characterize the star formation status of our sample using the specific star formation rate, defined as $\mathrm{sSFR} = \mathrm{SFR} / M_\star$, which provides a normalized metric of current star formation activity relative to the historical mass assembly of the galaxy. 
Our sample also exhibits a typical bimodal distribution in the $\mathrm{sSFR}$ versus $M_\star$ plane (Figure \ref{fig:M_SFR}). 
We adopt a threshold of $\log \mathrm{sSFR} / \text{yr}^{-1} = -11$ to distinguish between star-forming (active) and quenched (passive) populations\citep{Wetzel2012}. 
To decouple the confounding effects of stellar mass and local environment on galaxy evolution, we subdivide our mass-complete sample into three regimes: low-mass ($9.5 \le \log M_\star/M_\odot < 10$, 1569 galaxies), medium-mass ($10 \le \log M_\star/M_\odot < 10.5$, 1493 galaxies), and high-mass ($\log M_\star/M_\odot \ge 10.5$, 1488 galaxies).

\subsection{Local Surface Density ($\Sigma_5$)}

To quantitatively analyze how environmental density affects galaxy properties,
we adopt the $n$-th nearest-neighbor surface density estimator\citep{Schaefer2017}. The surface density is defined as: $\Sigma_n = n/(\pi r_n^2)$, where $r_n$ is the projected distance to the $n$-th nearest neighbor within the specified redshift slice. This technique effectively traces the underlying local matter density \citep{Muldrew2018}. In this work, we adopt the fifth nearest-neighbor density, $\Sigma_5$ ($n=5$), as our primary environmental metric. 
The distances between galaxies are calculated as projected distances and then converted to physical units ($\text{Mpc}$) based on the angular diameter distance at the redshift of A2029.

\begin{figure}
\centering
\includegraphics[width=0.7\textwidth]{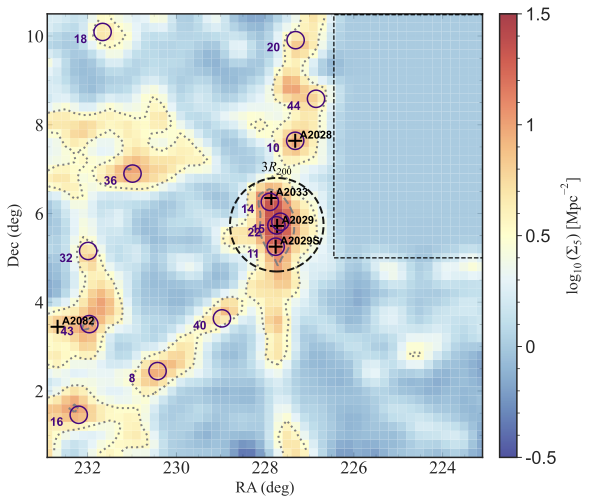}
\caption{Spatial distribution of local surface density ($\log_{10}\Sigma_5$) in the A2029 field. The dashed black contour delineates the $3R_{200}$ region of A2029. Black crosses mark the centroids of identified Abell clusters (A2029, A2028, A2033, and A2082). Grey dotted and dashed contours represent local galaxy surface density levels of $\log_{10}\Sigma_5 = 0.5$ and $1.0\ \mathrm{Mpc}^{-2}$, respectively. Purple circles denote the hierarchical substructures identified by the Blooming Tree (BT) algorithm \citep{Yu2025}.}
\label{fig:density}
\end{figure}

After calculating the $\Sigma_5$ for all galaxies, we examined their spatial distribution, as shown in Fig. \ref{fig:density}. In this 50$\times$50 mosaic, the color of each pixel represents the median $\Sigma_5$ value of all galaxies within that pixel, subsequently smoothed with a 2-pixels-wide Gaussian kernel. This map provides an approximate representation of the matter distribution within the field of view. The central massive region, including A2029, A2033 and A2029S($\log_{10} \Sigma_5 >1.0$), its surrounding groups and filaments(0.5$<\log_{10} \Sigma_5 <$1.0), and low density regions ($\log_{10} \Sigma_5<0.5$) are all clearly recognizable. 

To verify the physical reality of these structures, we compared them with those detected by the Blooming Tree (BT) algorithm in this field \citep{Yu2025}.
Within the specific redshift interval of the A2029 ($z = 0.06 \sim 0.09$), the BT algorithm identified 34 distinct structures. 
Among these structures, 14 systems satisfy the criteria of being sufficiently massive (velocity dispersion $> 250 \text{ km/s}$) and dense ($n/r_{50} > 2 \ \mathrm{Mpc}^{-1}$, where $n$ is the number of member galaxies and $r_{50}$ is the radius covering 50\% of members). Here, we have retained their ID from the original paper; for example, sub15 and sub22 correspond to A2029, sub14 represents A2033, and sub11 is A2029S.
Notably, all 14 structures are located within high-density regions.
The primary advantage of $\Sigma_5$ is that it provides a localized density environment for each galaxy, allowing for a precise comparison of environmental effects on galaxy evolution.

\section{Results} \label{sec:res}

\subsection{Quenched Fraction}

\begin{figure*}
\centering
\includegraphics[width=0.8\textwidth]{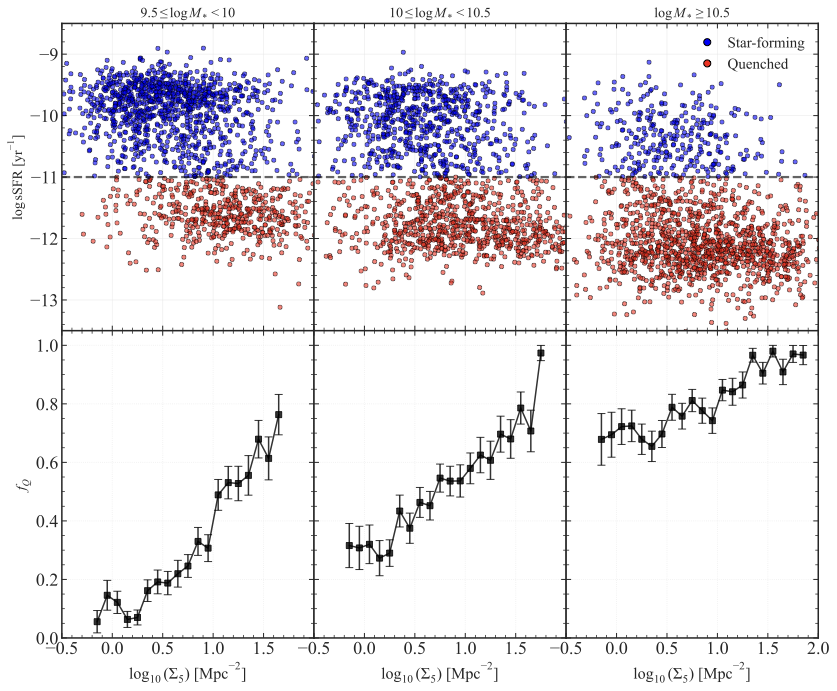}
\caption{Specific star formation rate (sSFR) and quenched fraction ($f_Q$) as a function of local surface density ($\Sigma_5$). The Top panels show the $\log \mathrm{sSFR}$ of galaxies versus their environment density for low-mass ($9.5 \leq \log M_\star/M_\odot < 10$; left), medium-mass($10 \leq \log M_\star/M_\odot < 10.5$; middle) and high-mass ($\log M_\star/M_\odot \geq 10.5$; right). The lower panels show the quenched fraction ($f_Q$) as a function of $\log_{10} \Sigma_5$ for the same stellar mass regimes. Error bars in the lower panels represent $1\sigma$ binomial uncertainties.}
\label{fig:sigma5_fq}
\end{figure*}

To quantify the joint impact of stellar mass and environmental density on galaxy quenching, we examine the quenched fraction ($f_Q$) across the three predefined mass regimes. Fig. \ref{fig:sigma5_fq} illustrates the variation of $f_Q$ as a function of the local surface density ($\log_{10} \Sigma_5$) for the low-mass, medium-mass, and high-mass subsamples.

For high-mass galaxies ($\log M_\star/M_\odot \ge 10.5$), the quenched fraction remains relatively high and stable ($f_q \approx 70\%$) in low-density environments ($\log_{10} \Sigma_5 < 0.5$). A gradual upward trend is observed within the intermediate density range of $0.5 < \log_{10} \Sigma_5 < 1.2$, beyond which star formation is suppressed in nearly all massive galaxies. This suggests that for massive galaxies, internal "mass quenching" establishes a high baseline for $f_Q$, which is further enhanced by environmental effects in groups and filaments before reaching a maximum in the cluster core \citep{Martig2009, Peng2010}.

In contrast, the $f_Q$ of medium-mass galaxies ($10.0 \le \log M_\star/M_\odot < 10.5$) in low-density regions is only approximately 30\%. 
This population exhibits a continuous and prominent increase in $f_Q$ across a broad density range ($0.5 \lesssim \log_{10} \Sigma_5 \lesssim 1.5$), approaching near-total quenching in the cluster core. 
The low-mass population ($9.5 \le \log M_\star/M_\odot < 10.0$) exhibits the most pronounced environmental response, characterized by the largest absolute variation in $f_Q$, which rises from $\sim 5\%$ in the field to $\sim 75\%$ in the cluster core.

\begin{figure*}
\centering
\includegraphics[width=0.45\textwidth]{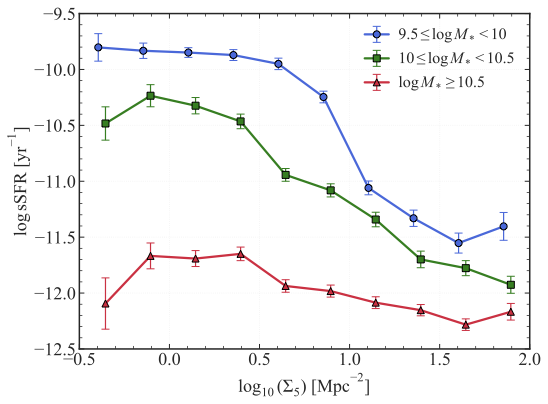}
\includegraphics[width=0.45\textwidth]{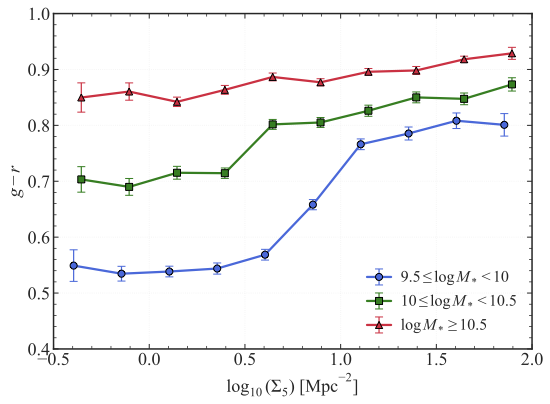}
\caption{Correlations between galaxy properties and local surface density ($\log_{10}\Sigma_5$). The panels illustrate the median trends of specific star formation rate ($\log \mathrm{sSFR}$; left) and $(g-r)$ color (right) as a function of environmental density. Different colored lines represent the three stellar mass subsamples, where the error bars denote the $1\sigma$ standard deviation.
}
\label{fig:propertiew}
\end{figure*}

These results indicate that the dynamic range of the star formation response to environmental density is strongly coupled with stellar mass.
To further compare the star formation activity across different mass regimes, we plot the median specific star formation rate ($ \log \mathrm{sSFR}$) as a function of local surface density ($\log_{10} \Sigma_5$) in the left panel of Fig. \ref{fig:propertiew}. All mass subsamples exhibit a systematic decline in $ \log \mathrm{sSFR}$ as the environment becomes progressively denser. 

For high-mass galaxies, the $\mathrm{sSFR}$ values are consistently suppressed across all environments, reflecting a state of quiescence driven primarily by internal evolutionary factors \citep{Bundy2006}.
Conversely, low- and medium-mass galaxies display a much more substantial absolute variation in their star formation activity,
maintaining relatively high $\mathrm{sSFR}$ in the field that drops precipitously upon approaching the cluster environment. 
This steep gradient of the low-mass population in the $\mathrm{sSFR} \text{--} \Sigma_5$ relation suggests that environmental mechanisms effectively truncate the cold gas supply on relatively short timescales upon accretion into higher-density nodes or filaments \citep{Wright2022}.

To isolate environmental quenching from internal mass-driven processes, we further analyze the Quenched Fraction Excess (QFE), following the methodology of \cite{Peng2010} and \cite{Zheng2024}. The QFE quantifies the fraction of galaxies that would have remained star-forming based on their mass alone but were subsequently quenched by their environment. It is defined as:
$$QFE = \frac{f_Q(\Sigma_5) - f_{Q, \text{field}}}{1 - f_{Q, \text{field}}}$$ 
where $f_{Q, \text{field}}$ is the quenched fraction in the lowest density bin ($\log_{10} \Sigma_5 < 0$), serving as the baseline for mass quenching.

\begin{figure*}
\centering
\includegraphics[width=0.45\textwidth]{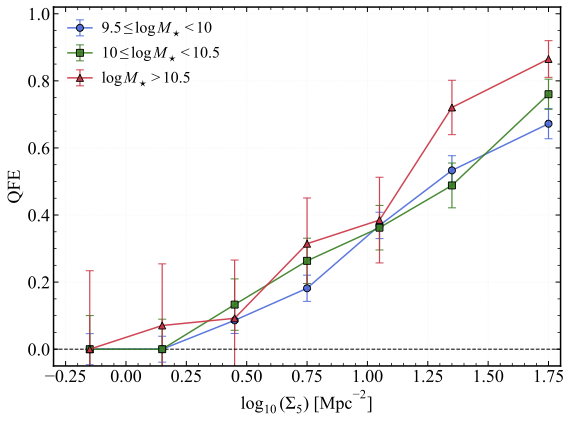}
\includegraphics[width=0.45\textwidth]{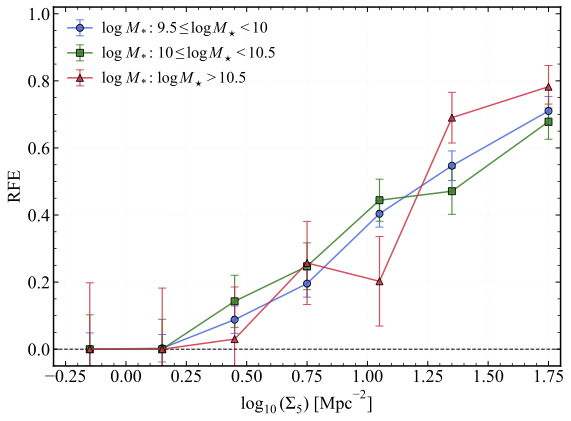}
\caption{Quenched Fraction Excess (QFE, left) and Red Fraction Excess (RFE, right) as a function of local surface density ($\log_{10} \Sigma_5$) for three stellar mass subsamples.In both panels, error bars represent $1\sigma$ binomial uncertainties, and the horizontal dashed lines at zero denote the field baseline established at $\log_{10} \Sigma_5 < 0$.}
\label{fig:qfe}
\end{figure*}

The resulting QFE as a function of $\log_{10}\Sigma_5$ is presented in the left panel of Figure~\ref{fig:qfe}.  Although the intrinsic quenched fraction varies among galaxies of different masses, they all exhibit consistent trends in low- to intermediate-density environments ($\Sigma_5 <$ 1.2).
While in high density regions ($\Sigma_5 >$ 1.2), the environmental effect is most pronounced in the high-mass galaxy population($\log_{10} M_\star/M_\odot \ge 10.5$). Almost all star-forming galaxies in this sample ceased star formation activity upon entering the dense environment.
In contrast, the proportion of low-mass galaxies that ceased star formation in the dense regions is not as high.
This result is consistent with previous study of \cite{Peng2010} and \cite{Zheng2024}.

\subsection{Red Fraction}
Compared to the star formation rate, color is a simple and direct observational quantity.
While the specific star formation rate ($\mathrm{sSFR}$) provides a detailed instantaneous measure of galactic activity, photometric colors offer a cheap and high-completeness proxy for long-term star formation history, as they are less susceptible to the uncertainties inherent in spectroscopic flux calibration or spectral synthesis modeling \citep{Strateva2001, Baldry2004}. Furthermore, the bimodal distribution of galaxy colors in the color-magnitude diagram has been extensively utilized to track integrated star formation histories, making it a powerful tool for large-sample studies across diverse cosmic environments \citep{Kauffmann2003, Blanton2003}. 

The trend in the median color as a function of density is shown in the right panel of Fig. \ref{fig:propertiew}. As environmental density increases, galaxies across all mass regimes exhibit a clear tendency to become redder, with the trend becoming more pronounced for lower-mass systems. This is consistent with our previous conclusions based on $\mathrm{sSFR}$ and aligns with other recent studies \citep{2024Nandi}.

\begin{figure*}
\centering
\includegraphics[width=0.8\textwidth]{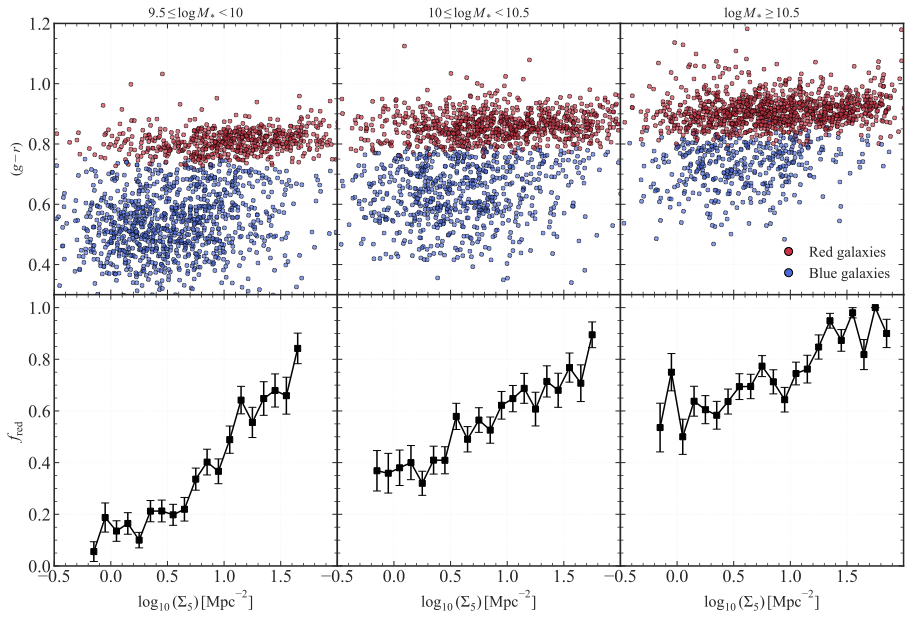}
\caption{Median $(g-r)$ color and red fraction ($f_{red}$) as a function of local surface density ($\Sigma_5$). Similar to Fig. \ref{fig:sigma5_fq}, the upper panels display the $(g-r)$ color versus $\log_{10} \Sigma_5$ for the low-mass (left), medium-mass (middle), and high-mass (right) subsamples. The lower panels illustrate the red fraction ($f_{red}$) as a function of $\log_{10} \Sigma_5$ across the same stellar mass regimes. }
\label{fig:sigma5_color}
\end{figure*}

To examine the environmental impact on galaxy color in detail, we investigate the relationship between the red galaxy fraction $f_{red}$ and the local surface density ($\log_{10} \Sigma_5$) across the three stellar mass subsamples, as shown in Fig. \ref{fig:sigma5_color}.
We find a significant positive correlation between the red fraction and local density for all mass populations, a trend that closely mirrors the behavior of the quenched fraction described in Section 3.1. This coherence reinforces the conclusion that high-density environments effectively facilitate the color transition from blue to red \citep{Strateva2001, Faber2007}.

For the high-mass population, $f_{red}$ remains consistently high(60\%) even in the low density regimes, suggesting that internal mass-driven processes, such as AGN feedback or morphological quenching, have already pushed these galaxies toward the red sequence prior to their accretion into denser structures \citep{Peng2010, Martig2009}. In contrast, 
the low-mass population exhibits a pronounced absolute variation in its red fraction. While the majority of low-mass galaxies in low-density regions are blue, their $f_{red}$ increases sharply as they approach the cluster environment ($\log_{10} \Sigma_5 \gtrsim 1.0$).  
This abrupt transition suggests that low-mass systems are particularly sensitive to environmental mechanisms, such as ram-pressure stripping or tidal harassment, which can rapidly remove gas reservoirs and induce color transformation on relatively short timescales \citep{Wetzel2013, Boselli2014}.

To further isolate the environmental drivers of color transformation, we introduce the red fraction excess (RFE), defined analogously to the QFE as
$$RFE = \frac{f_{\rm red}(\Sigma_5) - f_{\rm red, field}}{1 - f_{\rm red, field}}$$
where $f_{\rm red, field}$ is the red fraction in the lowest density bin ($\log_{10} \Sigma_5 < 0$). 
The RFE exhibits a trend similar to that of the QFE: in low- to intermediate-density environments, the trends in red fraction are consistent across different galaxy masses. While in high-density regions massive galaxies show a more complete transition to the red population.

The overall consistency between the $f_{red}$ and $f_Q$ trends indicates that both photometric color and star formation rate provide a unified picture of galaxy quenching within the A2029 field. Therefore, within a narrow range of redshifts, the color of a galaxy can be used as a rough indicator of its evolutionary stage.

\subsection{Color Map}

\begin{figure}
\centering
\includegraphics[width=0.7\textwidth]{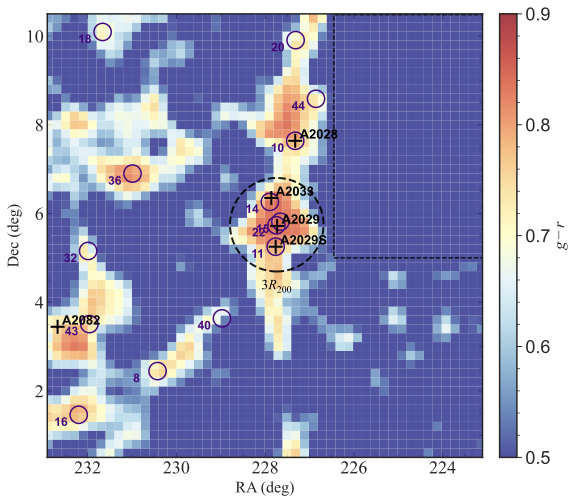}
\caption{Color density map of galaxies in the A2029 field. All symbols and contours are the same as in Fig. \ref{fig:density}.}
\label{fig:color}
\end{figure}

Since the color of galaxies is monotonically correlated with local matter density as the right panel of Fig. \ref{fig:propertiew} shows, its spatial distribution could also be used to reveal large-scale structures. Following the same procedure as for $\Sigma_5$, 
Figure \ref{fig:color} presents the spatial map of galaxy $(g-r)$ colors across the A2029 field, where each pixel represents the median color of all galaxies within that area.
The final 50$\times$50 pixel mosaic image was smoothed using a 2-pixel-wide Gaussian filter.

This color map is generally consistent with the $\log_{10}\Sigma_5$ density map and the 14 structures identified by the BT algorithm. 
However, the color of some structures (sub 32 and sub 40) exhibit colors that are not as red as anticipated. This may indicate that, although these regions have high galaxy densities, they are still in the early stages of assembly, and their member galaxies have not yet had enough time to consume their cold gas and complete the color transition.

\begin{figure}
\centering
\includegraphics[width=0.7\textwidth]{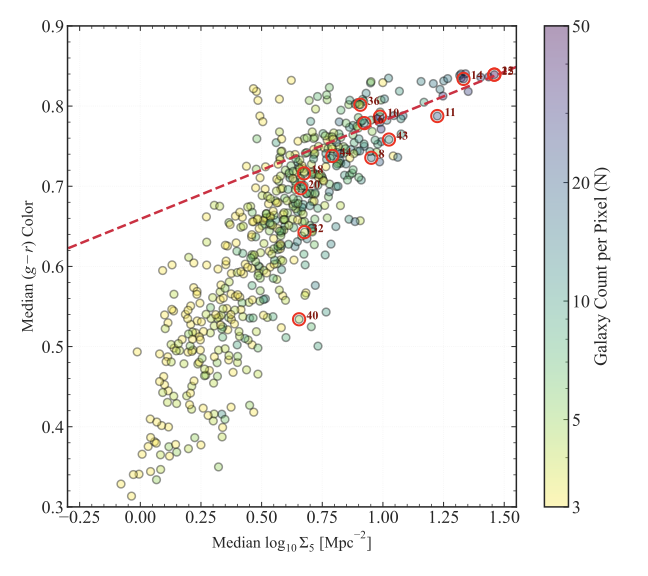}
\caption{Relationship between median $(g-r)$ color and local surface density $\log_{10} \Sigma_5$. Each grid pixel represents the median values within a spatial bin of approximately $12'$, color-coded by the number of galaxies ($N$) per bin. Red open circles denote the identified substructures from BT algorithm \citep{Yu2025}.
The red dashed line derived from the fit to the X-ray substructures. 
}
\label{fig:corr_all}
\end{figure}

To quantitatively compare the differences between the color map and the density map, we correlated the $\log_{10} \Sigma_5$ values with the $(g-r)$ colors for each pixel, as shown in Fig \ref{fig:corr_all}. A strong correlation is evident, implying that the large-scale structural features reflected in the color map are largely consistent with the results shown in the density map. In the regime where $\log_{10} \Sigma_5 < 1$, a nearly linear relationship between galaxy color and density. However, once the $\log_{10} \Sigma_5 >$ 1 and galaxies enter a cluster, the change in median color becomes gentile.

By overplotting the center properties of the 14 dense BT structures, we establish a reference line representing the color levels of different density structures. These virialized systems (except sub32 and sub40) are usually dominated by massive, red, and quenched galaxies.
Compared to them, galaxies in other regions at similar densities are mostly bluer.
One possible explanation is that their local density has been overestimated due to projection effects; another possibility is that these galaxies have not spent enough time in high-density environments to undergo color transformation.

\begin{figure*}
\centering
\includegraphics[width=0.3\textwidth]{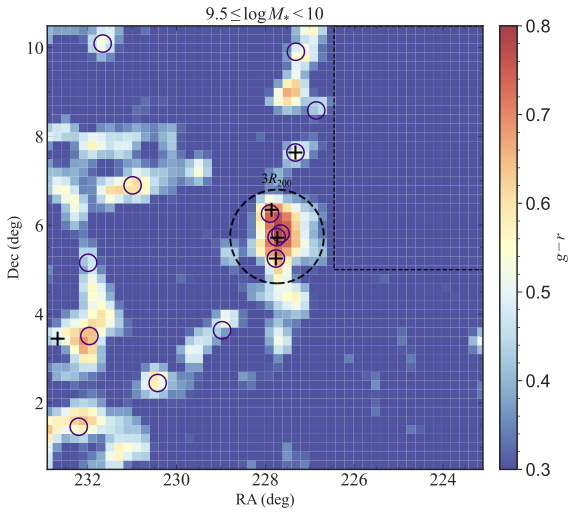}
\includegraphics[width=0.3\textwidth]{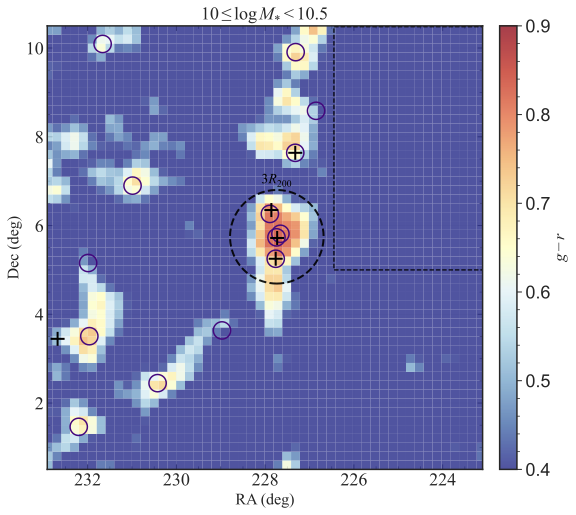}
\includegraphics[width=0.3\textwidth]{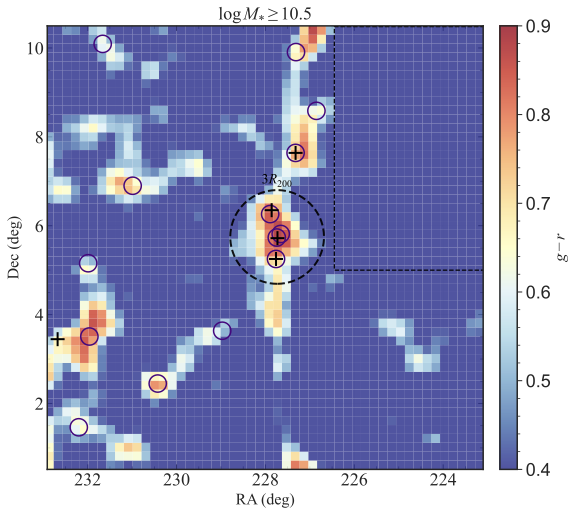}
\caption{Color density map of low-mass ($9.5 \leq \log M_\star/M_\odot < 10$; left), medium-mass ($10 \leq \log M_\star/M_\odot < 10.5$; middle) and high-mass ($\log M_\star/M_\odot \geq 10.5$; right). All symbols and contours are the same as in Fig. \ref{fig:density}.}
\label{fig:colormass}
\end{figure*}

\begin{figure}
\centering
\includegraphics[width=0.9\textwidth]{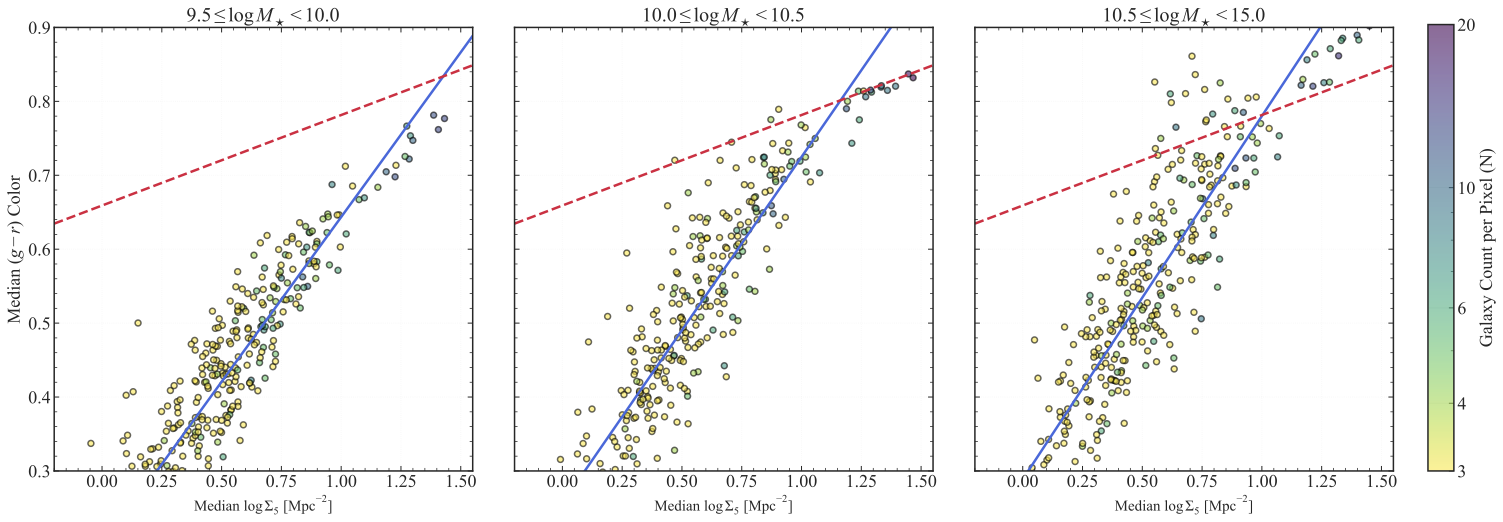}
\caption{Relationship between median $(g-r)$ color and local surface density $\log_{10} \Sigma_5$ across different stellar mass regimes. The panels from left to right represent the low-mass ($9.5 \leq \log M_\star/M_\odot  < 10.0$), medium-mass ($10.0 \leq \log M_\star/M_\odot  < 10.5$), and high-mass ($\log M_\star/M_\odot  \geq 10.5$).The blue solid lines show the best-fit relations for each mass regime. All other symbols and contours are the same as in Figure \ref{fig:corr_all}.
}
\label{fig:corr}
\end{figure}

Considering that massive galaxies can evolve toward the red sequence through intrinsic processes, while low-mass galaxies are primarily redden due to environmental factors, we generated the spatial color map of three mass subsamples separately (Fig. \ref{fig:colormass}), and then performed a correlation analysis between these plots and the density map at the pixel level (Fig. \ref{fig:corr}) to examine their differences.

We found that all three subsamples exhibit a strong correlation with environmental density, directly supporting the conclusion that the environment influences galaxy color. However, massive galaxies exhibit more scatter in their color-density relation. Its Pearson correlation coefficients ($r_p$) is 0.88. Even in regions of moderate density ($\log_{10} \Sigma_5 \sim$ 0.5 \text{--} 1), some pixels display very red colors ($g-r > $0.8). 
This can be explained by their powerful internal quenching mechanism.

For medium-mass galaxies, their data scatter is smaller than that of the massive subsample ($r_p=0.91$), and as the primary component of galaxy clusters, their color upper limit aligns very well with the reference line. Low-mass galaxies demonstrate the highest sensitivity to environmental density, exhibiting the strongest correlation ($r_p=0.92$). This tight correlation suggests that using the color distribution of low-mass galaxies to trace large-scale structures is a reasonable and effective approach.

\section{Conclusions} \label{sec:con}

In this study, we utilized highly complete spectroscopic data from the Dark Energy Spectroscopic Instrument (DESI) and the Sloan Digital Sky Survey (SDSS) to conduct a comprehensive environmental analysis of the A2029 field. By employing the local surface density ($\Sigma_5$) metric, we quantified the trends of the quenched fraction ($f_Q$) and red fraction ($f_\text{red}$) for three mass subsamples across a broad range of density regimes. Our primary conclusions are summarized as follows:

First, we find that both the $f_Q$ and the $f_\text{red}$ exhibit a statistically significant positive correlation with the local density across all stellar mass regimes. 
Notably, although the environmental dependence of these fractions (measured by QFE and RFE) is slightly weaker for low-mass galaxies than for their more massive counterparts, the absolute variation in their quenched and red populations remains remarkably large.

Second, we demonstrate that the color distribution of galaxies could serve as a promising tracer of the large-scale structure (LSS). While the colors of massive galaxies are influenced to some extent by intrinsic factors, the environmental impact acts as the primary driver for the color transition in low-mass galaxies.
Consequently, low-mass systems are well-suited for tracing variations in the local environment.

With the advancement of next-generation photometric surveys such as Eucild \citep{Euclid}, Large Synoptic Survey Telescope \citep[LSST;][]{LSST}, and Chinese Space Station Survey Telescope \citep[CSST;][]{CSST}, galaxy colors will serve as a valuable tool for studying large-scale structure.

\begin{acknowledgements}
This work has been supported by the National Natural Science Foundation of China No. 12573003, the National Key Research and Development Program of China (No. 2023YFC2206704), the China Manned Space Program with grant No. CMS-CSST-2025-A04 and the International Partnership Program of Chinese Academy of Sciences, Grant No.013GJHZ2024015FN. We thank the anonymous reviewer for the constructive comments and suggestions.

\end{acknowledgements}

\bibliographystyle{raa}
\bibliography{ms2026-0203}

\label{lastpage}

\end{document}